\documentclass[a4paper,10pt,twocolumn,twoside,journal]{article}

\usepackage{geometry}
\advance\oddsidemargin by -1in
\advance\evensidemargin by -1in
\advance\headsep by 2.8mm
\usepackage{cite}
\usepackage{hyperref}
\usepackage{url}
\usepackage{authblk}
\usepackage{mathptmx}
\usepackage{braket}
\usepackage{booktabs}
\usepackage{amsthm}
\usepackage{amssymb,amsmath,mdwlist,mathtools}
\usepackage{multirow}
\usepackage{multicol}
\usepackage{hhline}
\usepackage{algorithm}
\usepackage{algorithmicx}
\usepackage{algpseudocode}
\usepackage{graphicx}
\usepackage{caption}
\usepackage{subcaption}
\usepackage{adjustbox}
\usepackage{textcomp}
\usepackage{wrapfig}
\usepackage{tikz}
\usetikzlibrary{quantikz2}

\def\BibTeX{{\rm B\kern-.05em{\sc i\kern-.025em b}\kern-.08em
    T\kern-.1667em\lower.7ex\hbox{E}\kern-.125emX}}

\usepackage[normalem]{ulem}

\usepackage[compact]{titlesec}

\title{Optimized Compilation for Distributed Quantum Computing}

\author[1]{Michele~Bandini}
\author[1]{Davide~Ferrari}
\author[2]{Stefano~Carretta}
\author[1,*]{Michele~Amoretti}

\affil[1]{\small \textit{Quantum Software Laboratory}, Department of Engineering and Architecture, University of Parma, Parma, 43124 Italy (\href{https://www.qslab.unipr.it/}{https://www.qslab.unipr.it/})}
\affil[2]{\small Department of Mathematical, Physical and Computer Sciences, University of Parma, I-43124 Parma, Italy and INFN–Sezione di Milano-Bicocca, gruppo collegato di Parma, 43124 Parma, Italy}
\affil[*]{\small Corresponding author: Michele Amoretti, michele.amoretti@unipr.it.}

\date{}

\begin{document}

\maketitle

\begin{abstract}
\textbf{
In many practical applications, quantum algorithms require several qubits, significantly more than those available with current noisy intermediate-scale quantum processors. Distributed quantum computing (DQC) is considered a scalable approach to increasing the number of available qubits for computational tasks. In the DQC setting, a quantum compiler must find the best partitioning for the quantum algorithm and then perform smart non-local operations scheduling to optimize the consumption of Einstein-Podolsky-Rosen (EPR) pairs. In this work, the focus is on minimizing the use of EPR pairs when the circuit structure allows for multiple non-local gates to utilize a single TeleGate operation. This is achieved by using a greedy algorithm that explores the circuit and groups together the gates that could share an EPR pair while also changing the order of commutative gates when necessary. With this preliminary pass, the compiled circuits show reduced depth and EPR usage. Since the quality of each EPR pair quickly deteriorates, the number of non-local gates using the same EPR pair should also be bounded. This means that, depending on the features of the target quantum network, the user can achieve different levels of optimization. Here, it is shown that this approach brings benefits even while assuming a low EPR pair lifetime.
}
\end{abstract}


\maketitle

\section{Introduction}
\label{sec:introduction}
Most practical applications of quantum algorithms require many more qubits than those provided by current Noisy Intermediate-Scale Quantum (NISQ) platforms. Merely augmenting the number of physical qubits on a single device is not beneficial for the quality of computation because of the increasing noise~\cite{VanDev2016}. Future devices will adopt quantum error correction to extract a few high-quality logical qubits from many noisy physical qubits. Therefore, to supply users with many logical qubits, it could prove advantageous to exploit the Distributed Quantum Computing (DQC) paradigm \cite{DQCSurvey2024}, which involves splitting the circuits into subprograms that are distributed for execution over networked quantum processing units (QPUs).  

DQC efficiency and effectiveness will also depend on a well-designed quantum compiler, which is responsible for finding a suitable partitioning of the quantum algorithm and then appropriately scheduling non-local operations (i.e., two-qubit gates across different QPUs) to keep network resources to a minimum. Moreover, the quantum compiler must compute proper local transformations for each partition.

Compiling a quantum circuit is a very challenging task, even for a single-QPU architecture \cite{FerAmo2022}. Indeed, it is an NP-complete problem \cite{Botea2018} that relatively few papers have tackled in the DQC context. An upper bound on the overhead induced by quantum circuit compilation for DQC was found by Ferrari et al. \cite{FerCacAmo2021}. A common approach is to represent quantum circuits as edge-weighted graphs, with qubits as vertices. Edge weights correspond to the number of two-qubit gates \cite{Daei2020,Nikahd2021,Sundaram2021,SunGupRam2022}, reducing the problem to a minimum k-cut problem. Another idea is to represent circuits as bipartite graphs, with one set of vertices for the qubits and another for the gates, and to assign qubits to QPUs by means of a dynamic programming algorithm that seeks to minimize the number of non-local operations \cite{Dava2020}. In~\cite{CuoCalKrs2023}, the compilation problem is modeled as a generalization of the quickest multi-commodity flow problem and is solved by means of techniques from the literature. In~\cite{AndresMartinez2024}, the authors introduce several techniques for solving the problem, using Steiner trees to detect and reuse common connections, further reducing the cost of entanglement sharing within the network. In~\cite{Mengoni2025}, the authors propose a gate reordering approach.

In a previous work on this topic, Ferrari et al. \cite{Ferrari2023} introduced a modular compilation framework for DQC and presented the experimental evaluation of a compiler prototype. In that case, the authors considered network topologies with sparse connectivity, with EPR pairs (i.e., maximally entangled bipartite states) established between non-adjacent QPUs by making intermediate QPUs play the role of quantum repeaters -- an assumption that fits network-on-chip and local area network scenarios. 

In this paper, an improved framework is presented for a compiler that copes with all networking scenarios. 
The compiler can recognize when multiple non-local gates can be implemented with just one EPR pair, thus optimizing the consumption of network resources. The novelty of the proposed approach lies in a clever integration of non-local gate grouping, reordering, and scheduling, which results in the possibility of setting the optimization level. This is a common feature in compilers for classical computing, but it is rather new in quantum compilers. 

The paper is organized as follows. In Section \ref{sec:model}, the considered DQC system model is presented. In Section \ref{sec:compilation}, the compilation process is illustrated in detail. In Section \ref{sec:experiments}, the experimental evaluation of the compiler is presented and discussed. Finally, Section \ref{sec:conclusions} concludes the paper with an outline of future work.

\section{DQC System Model}
\label{sec:model}

In this section, the considered DQC system model is described, elucidating the main assumptions and the way non-local quantum gates can be implemented. 

\subsection{Assumptions}

The considered quantum computers are equipped with quantum communication capabilities, as well as Local Operations and Classical Communication (LOCC). Quantum connections between QPUs may be direct or mediated by quantum repeaters \cite{Talsma2024,ManAmo2022}. In both cases, the quantum communication channel can provide EPR pairs, which are assumed to be in the Bell state $\ket{\phi^+}$.

Since EPR pairs enable non-local gates, they are the primary resource for DQC. Therefore, in the proposed compiler, a DQC system is modeled as a graph $G =(V,E)$ such that each vertex $v \in V$ corresponds to a QPU, and each edge $(v_1,v_2) \in E$ indicates that EPR pairs may be shared between $v_1$ and $v_2$. Fully connected graphs match the vision of the Quantum Internet \cite{Wehner2018}. Sparse graphs, on the other hand, correspond to local area networks, either on-chip or room-scale.

\subsection{Implementation of Non-local Quantum Gates}
\label{sec:remotegates}

\begin{figure*}[!ht]
    \centering
    \begin{subfigure}{.49\linewidth}
        \includegraphics[width=.9\linewidth]{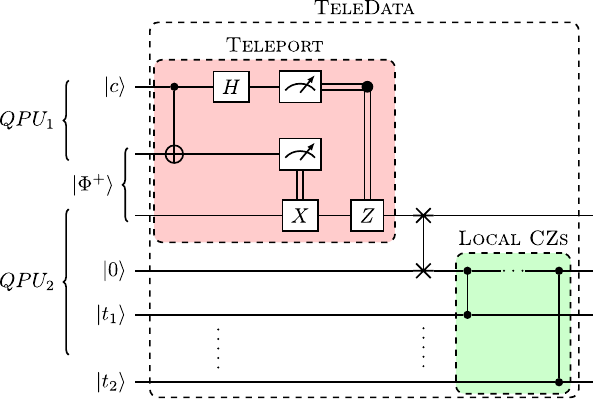}
        \caption{}
        \label{fig:tele-data}
    \end{subfigure}
    \hfil
    \begin{subfigure}{.49\linewidth}
        \includegraphics[width=.9\linewidth]{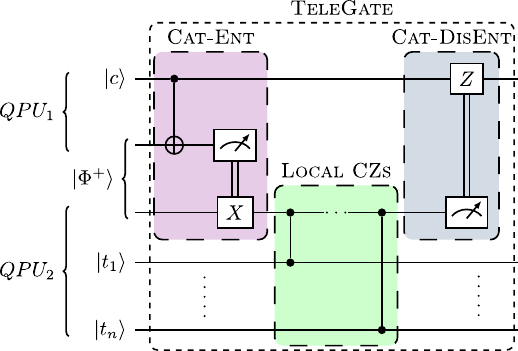}
        \caption{}
        \label{fig:tele-gate}
    \end{subfigure}
    \caption{(\subref{fig:tele-data}) Circuit representation of \texttt{TeleData} by means of the \textit{Teleport} primitive. \texttt{TeleData} moves the quantum state of a data qubit $\ket{c}$ to one qubit of an EPR pair and, then, swap it to a free data qubit. The state of the EPR pair and $\ket{c}$ are lost in the process. Multiple \texttt{CZ} acting on the teleported qubit can then be executed. (\subref{fig:tele-gate}) Circuit representation of \texttt{TeleGate} by means of \textit{Cat-Ent} and \textit{Cat-DisEnt} primitives. After the \textit{Cat-Ent} operation, the second qubit of the EPR pair participates in an entangled state with the control qubit. Multiple \texttt{CZ} with same control qubit and different target can be executed between \textit{Cat-Ent} and \textit{Cat-DisEnt}. It is worth noting that, between \textit{Cat-Ent} and \textit{Cat-DisEnt}, the control qubit is entangled with the EPR pair's one and cannot be targeted by other gates.}
    \label{fig:remote-ops}
    \hrulefill
\end{figure*}

Non-local gates can be realized using three communication primitives referred to as \textit{Teleport}~\cite{CacCalVan2020} (quantum state teleportation), \textit{Cat-Ent} (cat-entanglement), and \textit{Cat-DisEnt} (cat-disentanglement)~\cite{EJPP2003,Yimsiriwattana2005}. These primitives enable the execution of two distinct non-local operations, specifically \texttt{TeleData} and \texttt{TeleGate}~\cite{VanNem2006, VanMun2008, VanDev2016}, as shown in Figure~\ref{fig:remote-ops}.

\section{Compilation Process}
\label{sec:compilation}

\begin{figure*}[!ht]
    \centering
    \includegraphics[width=18cm]{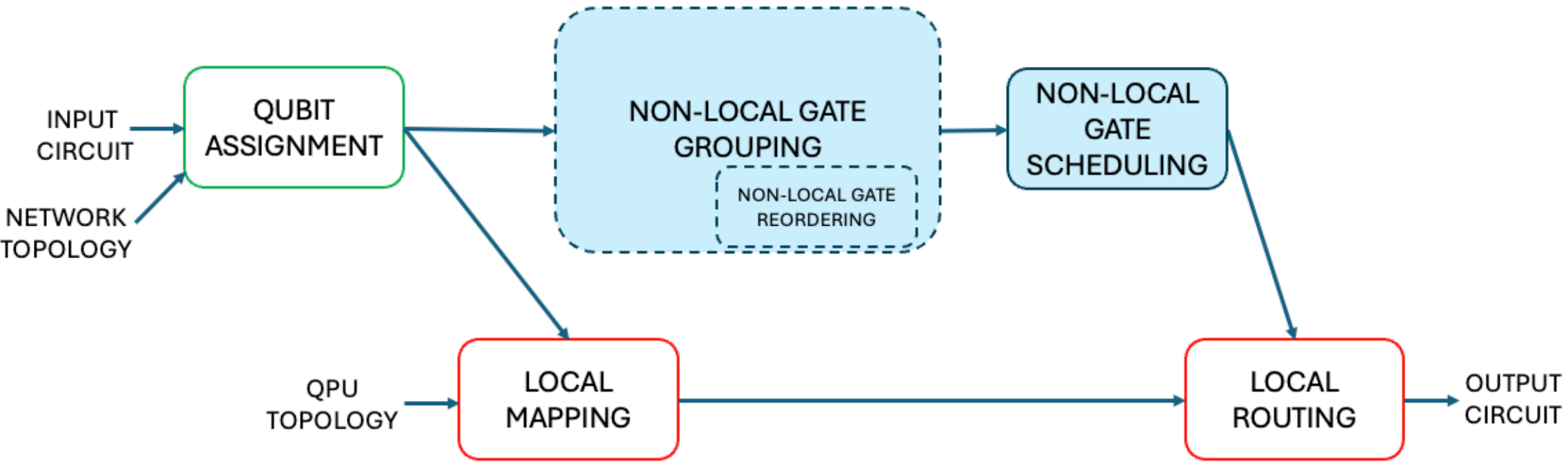}
    \caption{Flowchart describing the compilation process. The input includes an abstract description of a quantum circuit and a high-level description of the network configuration. The mandatory passes are qubit assignment, non-local gate scheduling, local mapping and local routing. Optional passes are non-local gate grouping and non-local gate reordering. The passes concerning non-local gates are outlined, as they are the novel contribution of this work.} 
    \label{fig:flowchart}
    \hrulefill
\end{figure*}

The compilation process is illustrated by the flowchart in Figure \ref{fig:flowchart}. The proposed compiler performs a number of \textit{passes}, with some degree of parallelism between them. The first one is qubit assignment, which deals with the partitioning of the input circuit. Then, the passes that deal with non-local gates (grouping, reordering, and scheduling) can be executed at the same time as the local mapping pass. The final pass is the local routing one. 

In situations of practical concern, the availability of entanglement is limited, and its usage is costly. This means that minimizing the number of required EPR pairs can produce important performance and cost benefits and can even impact the effective feasibility of some quantum circuits. To address this challenge, non-local gate grouping and non-local gate reordering have been included as optional passes in the compiler. Their activation within the compilation process enables optimized output circuits but requires a specific non-local gate scheduling strategy. 

The three passes concerning non-local gates are the main contributions of this work. Their detailed descriptions are provided in the following subsections. 

The local mapping and routing passes were thoroughly presented in previous works~\cite{Ferrari2018,Ferrari2021,Ferrari2023}. Therefore, in this paper, they are not described in detail, nor are they included in the performance evaluation (Section \ref{sec:experiments}). The local mapping pass associates the qubits of each sub-circuit with the qubits of the QPU that will execute that sub-circuit. The local routing pass scans each sub-circuit and, for every gate that has to be executed across qubits of the QPU that are not directly connected, according to the coupling map of the device, computes the shortest sequence of necessary SWAP gates. 

The current implementation of the compiler is available on GitHub \cite{dqc-nac}. Input circuits are described using OpenQASM~\cite{OpenQASM}, while output circuits are represented using Qoala~\cite{Qoala2025}, a unified program format for hybrid interactive classical-quantum programs. This means that the compiler's output is ready to be executed on real DQC platforms whose nodes support the Qoala language.

\subsection{Qubit Assignment}
\label{subsec:qubit_assignment}

The goal of the qubit assignment pass is to partition the circuit in order to minimize the communication cost, i.e., the number of consumed EPR pairs. The approach used here is similar to the one presented in \cite{Ferrari2023}, with one key difference: since grouping techniques may be employed later to utilize one EPR pair for more than one non-local operation, the goal cannot simply be to minimize the number of non-local gates, but also to create a situation where the grouping strategy described in Section \ref{subsec:grouping} works better.

As a premise, the current qubit assignment pass tries to split the input circuit by assigning the same number of qubits to each QPU. That is, if the input circuit has $n_q$ qubits and the target DQC system has $N$ QPUs, then $n_q/N$ qubits will be tentatively assigned to each QPU.

An undirected weighted graph $G_c(V_c,E_c)$ is used to represent the circuit $c$, where each vertex $v\in V_c$ represents a qubit in $c$, and the weight $W(e)$, for each $e\in E_c$, represents the number of EPR pairs needed to connect the two qubits in case they are in two different QPUs. Then the problem simplifies to one of graph partitioning, where the objective is to compute a k-way partitioning such that the sum of inter-partition edge weights is minimized. Here, the same strategy adopted in \cite{Ferrari2023} is used.

\begin{figure}[!ht]
     \centering
     \begin{subfigure}{.4\textwidth}
        \centering
        \includegraphics[width=1.\textwidth]{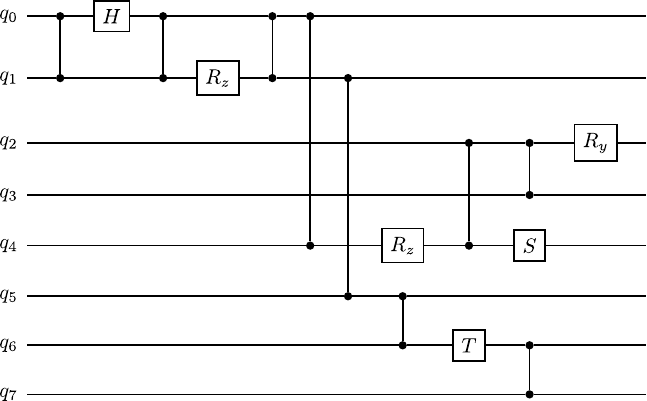}
        \caption{}
        \label{fig:partition_start}
     \end{subfigure}
     
     \begin{subfigure}{0.4\textwidth}
        \centering
        \includegraphics[width=1.\textwidth]{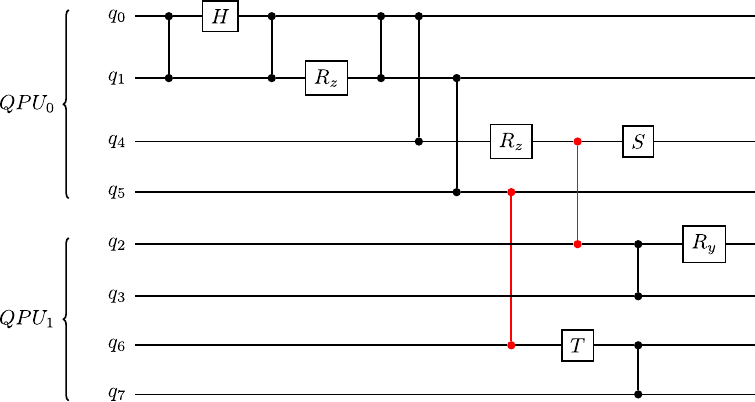}
        \caption{}
        \label{fig:partition_normal}
     \end{subfigure}
    
     \begin{subfigure}{0.4\textwidth}
        \centering
        \includegraphics[width=1.\textwidth]{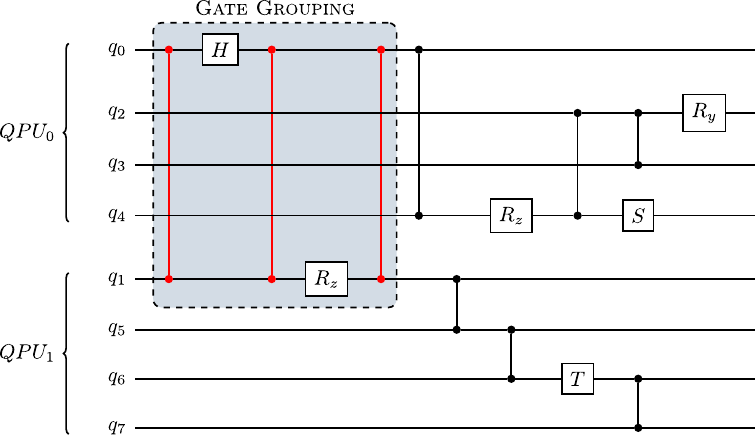}
        \caption{}
        \label{fig:partition_pre_pass}
     \end{subfigure}

    \caption{Example of qubit assignment. \textbf{(a)} Input circuit. \textbf{(b)} Baseline qubit assignment as in \cite{Ferrari2023}. \textbf{(c)} Qubit assignment with non-local gate grouping.}
    \label{fig:qubit_assign}
    \hrulefill
\end{figure}

The harder task is to properly count the number of EPR pairs that are needed. As can be seen in Figure \ref{fig:qubit_assign}, if the count was done on the number of two-qubit gates, the partition computed as the best one would need two EPR pairs to execute all the non-local gates, as shown in Figure \ref{fig:partition_normal}. With the grouping strategies described in Section \ref{subsec:grouping}, the result is that only one EPR pair is used, as shown in Figure \ref{fig:partition_pre_pass}. Here, the three $CZ$ gates are considered to have $q_1$ as their control and $q_0$ as their target, and therefore the $H$ gate does not cause problems since it acts on the target. The $R_z$ can commute, as per the rules stated in Section \ref{subsec:gate_reordering}. Unlike the grouping strategy described in Section \ref{subsec:grouping}, two-qubit gates that only share one qubit are always considered to interfere with each other because a proper check can only be done with the qubit mapping finalized.

\subsection{Non-local Gate Grouping}
\label{subsec:grouping}

This optimization pass works in a greedy manner by grouping non-interfering gates into a single multi-qubit instruction before the compilation phase. The pseudo-code in Algorithm~\ref{alg:gate-grouping} describes the optimization process. It accepts a gate array $GA$ as input and, as it progresses through the process, creates an output circuit $OC$ that contains single or grouped gates. There is a limit $D_{max}$ to the number of consecutive gates that can be included in a group. The reasons for this are better explained in Section~\ref{subsubsec:ent-time-limit}.

\subsubsection{Algorithm Breakdown}
\label{subsubsec:alg-breakdown}

For each qubit, a working memory is maintained, which initially points to nothing but could later point to a gate group that is created during the search for non-local gates. All the gates in $GA$ are scanned in order. Every time the optimization algorithm finds a new gate that has no qubit present in any qubit's working memory, it checks if it is local; in that case, it simply puts it into $OC$; otherwise, it creates a new group that contains only the gate and makes every memory relative to the qubits of the gate point towards the group.

If the gate instead has some qubits in a working memory, either its own or another, the gate-grouping pass checks if they all point to the same group; if they do, it checks whether the gate interferes. If it does not interfere, the gate-grouping pass adds it to the group in question. This triggers a control that checks if the group has reached the maximum number of gates per EPR pair $D_{max}$: if it has, the group is added to $OC$, and the memory is released.

If the gate interferes with the group, it checks if it can commute with the next gate; in that case, it repeats the process with the next gate in $GA$. If the gate interferes with a group and cannot be solved by reordering, the optimization module simply adds all the gates in the group to $OC$, releases the memory, and repeats the process again. In either case, every time a gate is placed in a group or directly in $OC$, it is removed from $GA$, and the optimization process starts again by scanning the first (remaining) gate.

\begin{algorithm}[!ht]
    \caption{
        \newline
        \footnotesize
        \textbf{Input}: Array of gates $GA$, network layout $N$, number of qubits $m$, maximum duration of EPR pair $D_{max}$
        \newline
        \textbf{Output}: circuit $OC$ with grouped-together gates
    }
    \label{alg:gate-grouping}
    \begin{algorithmic}[1]
        \vspace{0.2cm}
        \Function{Gate-grouping}{$GA, N, m, D_{max}$}
            \State $OC \gets Null$
            \State $M \gets Null$
            \For{$i$ from $1$ to $m$}
                \State $M[i] \gets Null$
            \EndFor
            \State $j \gets 0$
            \While{$length(GA) > 0$} \label{while}
                \State \Call{Check-gate}{$j$, $OC$, $M$, $D_{max}$}
            \EndWhile     
        \EndFunction
        \vspace{0.2cm}
        \Function{Check-gate}{$j$, $OC$, $M$, $D_{max}$}
            \State $g \gets GA[j]$
            \If{$\forall i \in g.qubits, M[i] == Null$}
                \If{g is local}
                    \State \textbf{append} $g$ to $OC$
                \Else
                    \State $G \gets [g]$
                    \ForAll{$i \in g.qubits$}
                        \State M[i] points to G
                    \EndFor
                \EndIf
            \ElsIf{$\forall i \in g.qubits, M[i]$ points to the same Group, where $g$ can be added}
                \State $G \gets$ the Group which $M[i]$ points, $\forall i$
                \State \textbf{append} $g$ into G
            \ElsIf{GA[j+1] can commute with g}
                \State $j \gets j+1$
                \State \Return
            \Else
                \State $G \gets$ the first Group s.t. $\exists i$ s.t. $M[i]$ points to it
                \State \textbf{append} $G$ into $OC$ 
                \State $\forall i \in g.qubits, M[i] \gets Null$
                \State \Return 
            \EndIf
            \If {$G.length == D_{max}$}
                \State \textbf{append} $G$ into $OC$ 
                \State $\forall i \in g.qubits, M[i] \gets Null$
            \EndIf
            \State delete $GA[j]$
            \State \Return
            
        \EndFunction
    \end{algorithmic}
\end{algorithm}

\subsubsection{Optimizations when Entanglement Duration is Limited}
\label{subsubsec:ent-time-limit}
So far, in the literature on compilers for DQC, the approach has always been to maximize the number of non-local quantum gates that can be implemented with a single EPR pair. As anticipated above, the non-local gate grouping pass proposed in this work defines a settable limit $D_{max}$ on the maximum depth of a group, which is a fair representation of the time spent keeping the same EPR pair in use.

This approach is realistic, as it is difficult to keep an entangled pair for a very long time in practical contexts, particularly over the Quantum Internet. Of course, this reduces the performance gains from the gate grouping, but as shown in Section~\ref{sec:experiments}, the results are still positive even with a very small $D_{max}$. This makes the proposed algorithm suitable for real applications.

\subsection{Non-Local Gate Reordering}
\label{subsec:gate_reordering}

As shown in Figure \ref{fig:qubit_assign}, sometimes, to better group the non-local gates and minimize EPR usage, gate commutativity can be capitalized on to achieve improved results. This is better done after the qubit assignment (described in Section \ref{subsec:qubit_assignment}) has been completed. Therefore, the non-local gate reordering strategy is used inside the non-local gate grouping pass (lines 26--29 in Algorithm \ref{alg:gate-grouping}).

In general, when compiling, the objective is to obtain an equivalent circuit. Sometimes, gates can be applied in different orders without changing the resulting unitary. This occurs when the matrices representing the gates share an eigenbasis. For a lengthy circuit with many 2-qubit gates, there may be a need to make many computations, which could add up to be costly. 

However, when compiling, only a few basis gates are used, as they form a complete set. This means that, knowing the computational basis, some rules can be defined for the considered gates and applied to verify whether two gates can be commuted.

The main rules adopted by this pass are listed below. 
\begin{itemize}
    \item An $R_x$ gate commutes with a $CX$ gate if the rotation is applied to the target qubit
    \item An $R_z$ gate commutes with a $CX$ gate if the rotation is applied to the control qubit
    \item An $R_z$ gate always commutes with a $CZ$ gate
    \item Two consecutive $CX$ gates commute if their control qubit is the same
    \item Two consecutive $CX$ gates commute if their target qubit is the same
    \item Two consecutive $CZ$ gates always commute
\end{itemize}

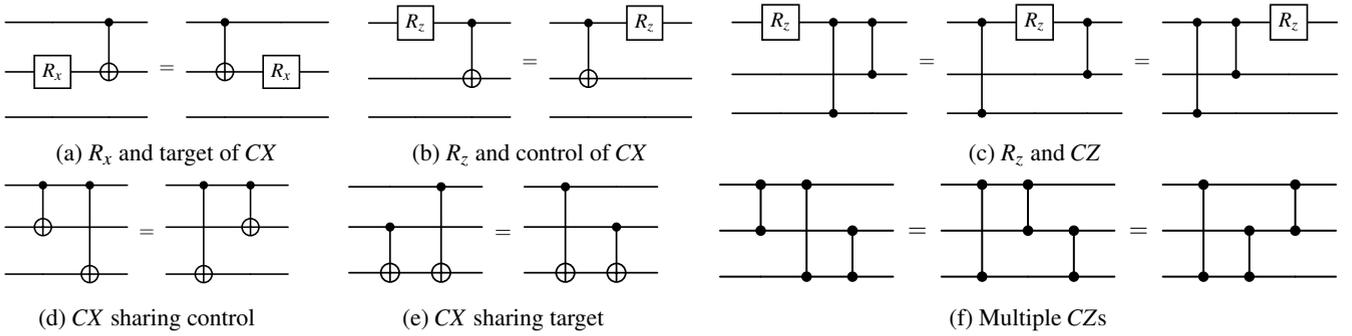
\begin{figure*}[!ht]
    \centering
    \begin{subfigure}{0.25\textwidth}
        \centering
        \begin{adjustbox}{width=1.\linewidth}
            \begin{quantikz}
            &  & \ctrl{1} & \\
            & \gate{R_x}&\targ{}& \\
            &&&
            \end{quantikz} $=$
            \begin{quantikz}
            & \ctrl{1}  &  & \\
            & \targ{}& \gate{R_x}& \\
            &&&
            \end{quantikz}
        \end{adjustbox}
        \caption{$R_x$ and target of $CX$}
        \label{fig:Rx_and_CX}
    \end{subfigure}
    \hfill
    \begin{subfigure}{.25\textwidth}
        \centering
        \begin{adjustbox}{width=1.\linewidth}
            \begin{quantikz}
            & \gate{R_z} & \ctrl{1} & \\
            &&\targ{}& \\
            &&&
            \end{quantikz} $=$
            \begin{quantikz}
             & \ctrl{1}  & \gate{R_z} & \\
             & \targ{}& & \\
             &&&
            \end{quantikz}
        \end{adjustbox}
        \caption{$R_z$ and control of $CX$}
        \label{fig:Rz_and_CX}
    \end{subfigure}
    \hfill
    \begin{subfigure}{0.46\textwidth}
        \centering
        \begin{adjustbox}{width=1.\linewidth}
            \begin{quantikz}
            & \gate{R_z} & \ctrl{2} & \ctrl{1} & \\
            &  && \ctrl{0} & \\
            && \ctrl{0} &  &
            \end{quantikz} $=$
            \begin{quantikz}
            & \ctrl{2} & \gate{R_z} & \ctrl{1} & \\
            &&  & \ctrl{0} & \\
            &\ctrl{0}&&  &
            \end{quantikz} $=$
            \begin{quantikz}
            & \ctrl{2} & \ctrl{1} & \gate{R_z} & \\
            && \ctrl{0} &  & \\
            &\ctrl{0}&  &  &
            \end{quantikz}
        \end{adjustbox}
        \caption{$R_z$ and $CZ$}
        \label{fig:Rz_and_CZ}
    \end{subfigure}
    \\
    \begin{subfigure}{0.22\textwidth}
        \centering
        \begin{adjustbox}{width=1.\linewidth}
            \begin{quantikz}
            & \ctrl{1} & \ctrl{2} & \\
            & \targ{} & & \\
            &&\targ{}&
            \end{quantikz} $=$
            \begin{quantikz}
            & \ctrl{2} & \ctrl{1} & \\
            && \targ{} & \\
            &\targ{}& &
            \end{quantikz}
        \end{adjustbox}
        \caption{$CX$ sharing control}
        \label{fig:Cx_control}
    \end{subfigure}
    \hfill
    \begin{subfigure}{0.24\textwidth}
        \centering
            \begin{adjustbox}{width=1.\linewidth}
            \begin{quantikz}
            &  & \ctrl{2} & \\
            & \ctrl{1}& & \\
            &\targ{}&\targ{}&
            \end{quantikz} $=$
            \begin{quantikz}
            & \ctrl{2}  & & \\
            & & \ctrl{1}& \\
            &\targ{}&\targ{}&
            \end{quantikz}
        \end{adjustbox}
        \caption{$CX$ sharing target}
        \label{fig:CX_target}
    \end{subfigure}
    \hfill
    \begin{subfigure}{0.47\textwidth}
        \centering
        \begin{adjustbox}{width=1.\linewidth}
            \begin{quantikz}
            & \ctrl{1} & \ctrl{2} &  & \\
            & \ctrl{0} && \ctrl{1} & \\
            &&\ctrl{0}& \ctrl{0} &
            \end{quantikz} $=$
            \begin{quantikz}
            & \ctrl{2} & \ctrl{1} &  & \\
            && \ctrl{0} & \ctrl{1} & \\
            &\ctrl{0}&& \ctrl{0} &
            \end{quantikz} $=$
            \begin{quantikz}
            & \ctrl{2} &  & \ctrl{1} & \\
            && \ctrl{1} & \ctrl{0} & \\
            &\ctrl{0}& \ctrl{0} &  &
            \end{quantikz}
        \end{adjustbox}
        \caption{Multiple $CZ$s}
        \label{fig:Multiple_CZs}
     \end{subfigure}

    \caption{Commutation rules}
    \label{fig:commutation_rules}
    \hrulefill
\end{figure*}

\subsection{Non-local Gate Scheduling}
Gate scheduling is performed in a simple way after gate grouping (described in Section \ref{subsec:grouping}) has been completed. Every non-local gate is inside a group, even if the group includes only one non-local gate. Therefore, every other gate is local and can be scheduled directly, as explained in \cite{Ferrari2023}.

When a group is reached, a \textit{Cat-Ent} is inserted. Then, every gate in the group is compiled in the same manner as described in \cite{Ferrari2023}. At the end of each group, a \textit{Cat-DisEnt} is inserted, completing the \texttt{TeleGate} primitive. If the network topology is complete, there is always a direct connection between the two QPUs. Otherwise, the shortest path between them has to be computed, and intermediate nodes are used as quantum repeaters.

\section{Experimental Evaluation}
\label{sec:experiments}

The compiler has been tested using circuits from the QASMBench \cite{QASMBench} benchmark suite, an OpenQASM benchmark suite for NISQ evaluation. Each circuit is characterized by the number of qubits $n_q$ and depth. The network topologies considered are complete, which means that each quantum node in the DQC system can share EPR pairs with any other node. Quantum nodes are characterized by the number of data qubits $n_d$ and the number of communication qubits $n_c$. Since the experimental evaluation regarded only non-local gate related passes, the characterization of the internal connectivity (coupling map) of the quantum nodes is not relevant.

In Table \ref{tab:results}, compilation results are reported for a large set of practical circuits, considering two quantum computers equipped with enough data qubits (i.e., $n_q/2 \leq n_d < n_q$) and four communication qubits (i.e., $n_c = 4$). The performance metric considered is the required number of EPR pairs. Three compiler configurations are compared: without optimization (i.e., no gate grouping, no gate reordering), with limited optimization (i.e., grouping non-local gates whose execution fits a limited time slot with gate reordering), and with unlimited optimization (i.e., grouping as many non-local gates as possible with gate reordering). It is evident that the compiler with optimization significantly outperforms the compiler without optimization \cite{Ferrari2023}.
These results are comparable to those of \cite{Mengoni2025}, with the disclaimer that the source code and experimental details of that paper are not available at this time.

\begin{table}[!ht]
    \centering
    \begin{tabular}{cc|ccc}
        \hline
         &  & & EPR pairs  & \\
        \hline
        Circuit & $n_q$ & no opt & limit=3 opt & unlimited opt \\
        \hline\hline
        Adder & 28 & 11 & 7 & 5 \\ \hline
        Adder & 64 & 11 & 7 & 5 \\ \hline
        Adder & 118 & 11 & 7 & 5 \\ \hline
        Cat & 35 & 1 & 1 & 1 \\ \hline
        DNN & 33 & 36 & 20 & 6 \\ \hline
        DNN & 51 & 52 & 28 & 6 \\ \hline
        GHZ & 40 & 1 & 1 & 1 \\ \hline
        Ising & 34 & 2 & 2 & 2 \\ \hline
        KNN & 41 & 36 & 18 & 6 \\ \hline
        Multiplier & 45 & 290 & 209 & 159 \\ \hline
        Multiplier & 75 & 862 & 604 & 459 \\ \hline
        Multiplier & 350 & 13228 & 11108 & 8761 \\ \hline
        QFT & 29 & 410 & 230 & 118 \\ \hline
        QFT & 63 & 702 & 364 & 51 \\ \hline
        QFT & 320 & 702 & 364 & 51 \\ \hline
        QuGAN & 39 & 40 & 22 & 6 \\ \hline
        QuGAN & 71 & 72 & 45 & 13 \\ \hline
        QuGAN & 111 & 112 & 66 & 14 \\ \hline
        QuGAN & 395 & 408 & 319 & 29 \\ \hline
        QV & 32 & 589 & 589 & 585 \\ \hline
        QV & 100 & 6526 & 6522 & 6442 \\ \hline
        Swap Test & 25 & 24 & 12 & 4 \\ \hline
        Swap Test & 41 & 36 & 18 & 6 \\ \hline
        Swap Test & 83 & 80 & 40 & 4 \\ \hline
        W-State & 76 & 2 & 2 & 2 \\ \hline
        W-State & 118 & 2 & 2 & 2 \\ \hline
        Square root & 45 & 5415 & 4411 & 3983 \\ \hline
    \end{tabular}
    \caption{The name and the number of qubits in the input circuit are in the first two columns, followed by the number of non-local gates  (corresponding to the baseline number of EPR pairs when no optimization is applied), then the number of EPR pairs when limiting the EPR lifetime to 3 time slots, and finally when optimizing without limitations.}
    \label{tab:results}
\end{table}

\begin{figure*}
     \centering
     \begin{subfigure}{.45\textwidth}
        \centering
        \caption{}
        \includegraphics[width=1.\textwidth]{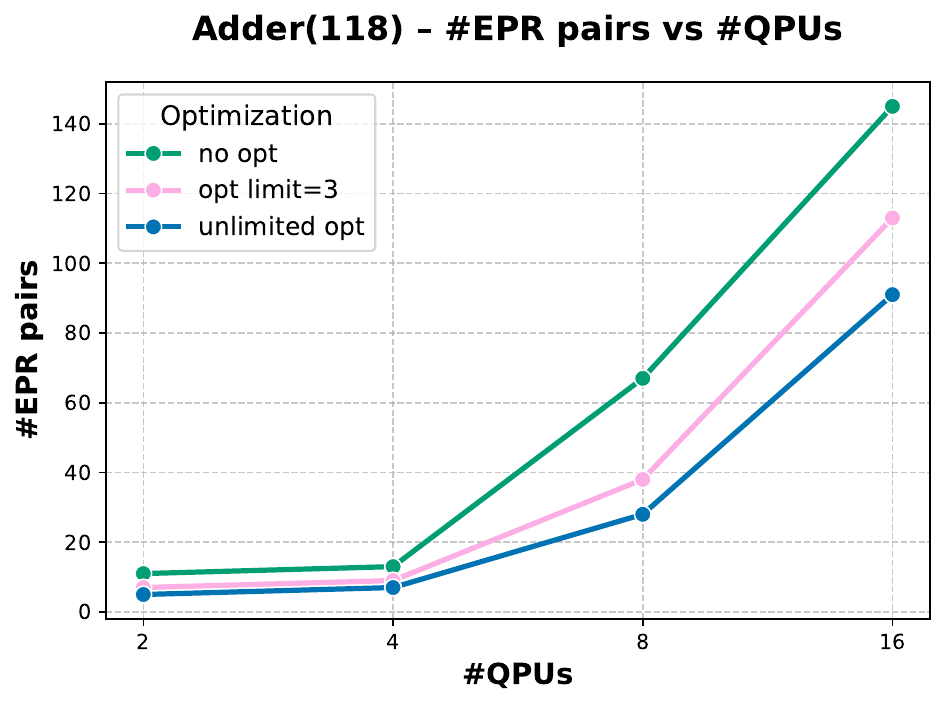}
        \label{fig:Adder_EPR_QPU}
     \end{subfigure}
     \hfill
     \begin{subfigure}{0.45\textwidth}
        \centering
        \caption{}
        \includegraphics[width=1.\textwidth]{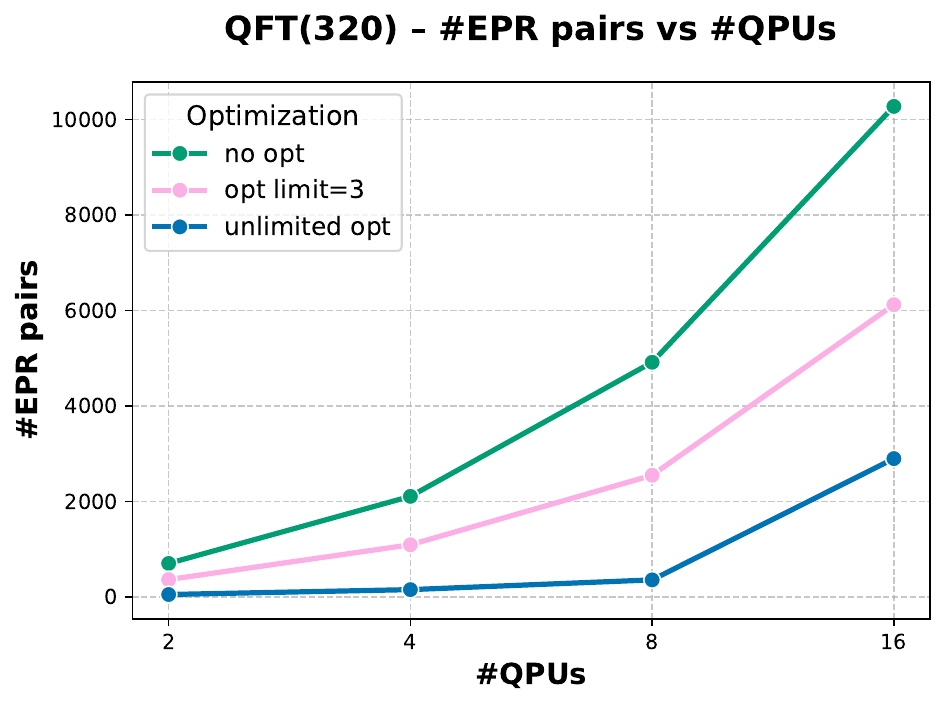}
        \label{fig:QFT_EPR_QPU}
     \end{subfigure}
     \\
     \begin{subfigure}{0.45\textwidth}
        \centering
        \caption{}
        \includegraphics[width=1.\textwidth]{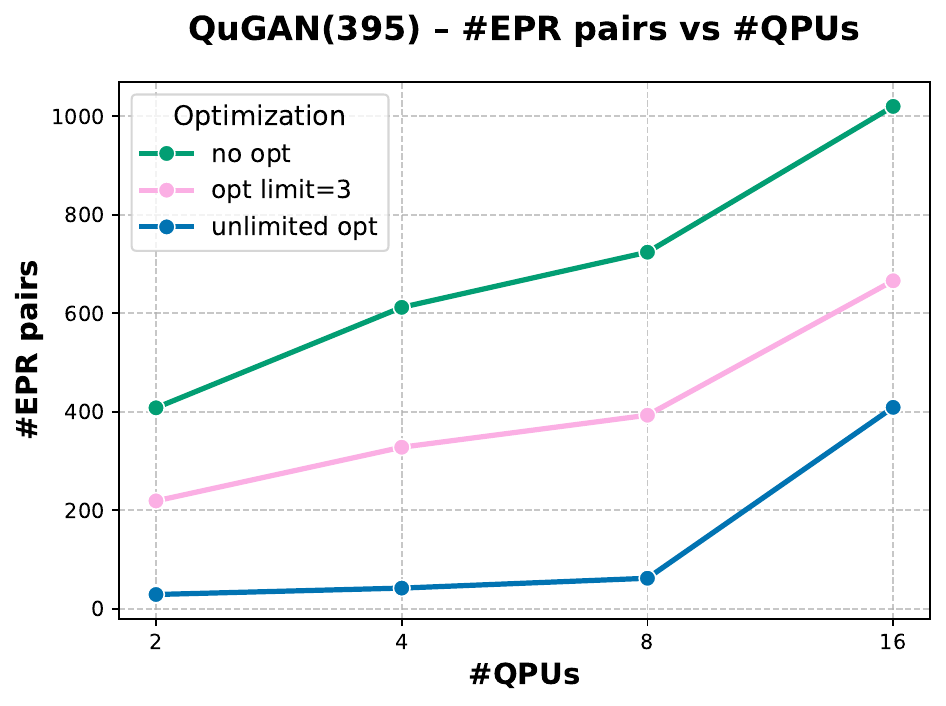}
        \label{fig:QuGAN_EPR_QPU}
     \end{subfigure}
     \hfill
     \begin{subfigure}{0.45\textwidth}
        \centering
        \caption{}
        \includegraphics[width=1.\textwidth]{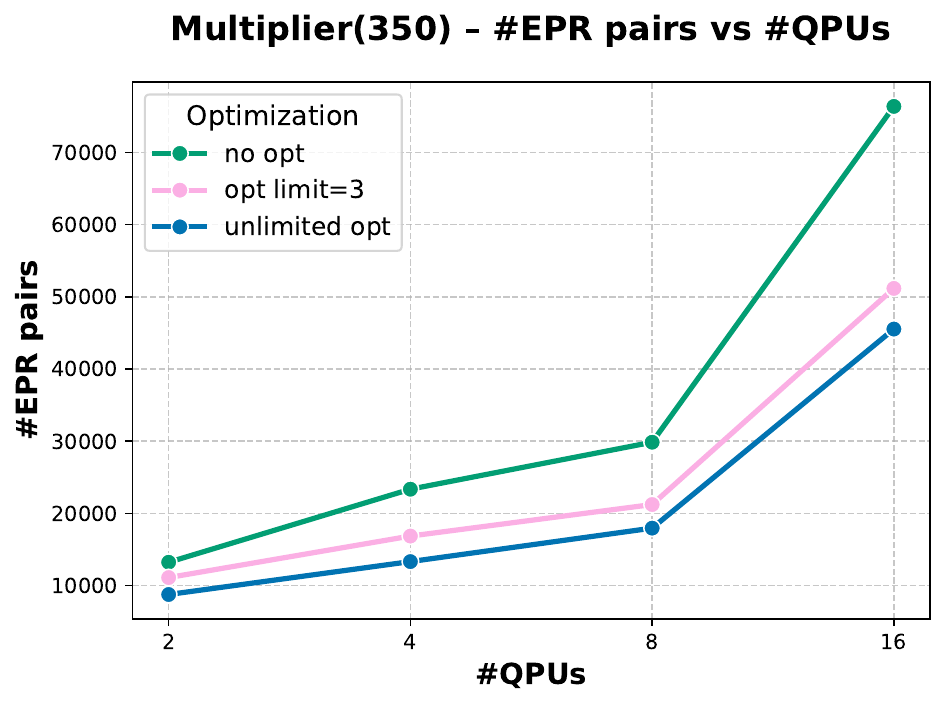}
        \label{fig:Multiplier_EPR_QPU}
     \end{subfigure}
     \hfill
     
    \caption{Number of required EPR pairs with respect to the number of QPUs, for selected quantum circuits with very different features. In each plot, different compiler configurations are compared. For (a) and (b), the four considered DQC architectures include devices with 64, 32, 16 and 8 data qubits, respectively. For (c) and (d), instead, the four considered DQC architectures include devices with 256, 128, 64 and 32 qubits, respectively. Each device has 4 communication qubits.}
    \label{fig:EPR_vs_QPU}
    \hrulefill
\end{figure*}

For selected circuits, Figure \ref{fig:EPR_vs_QPU} shows how many EPR pairs are used when the circuits are compiled for different DQC architectures. Specifically, from one architecture to another, the number of QPU grows exponentially, while the number of qubits per QPU decreases accordingly. In each plot, three compiler configurations are compared: without optimization, with unlimited optimization, and with limited optimization. It may be observed that, as expected, the number of requested EPR pairs grows with the number of QPUs for all three compiler configurations. Importantly, the growth is less steep when optimization is used. 

\begin{figure*}
     \centering
     \begin{subfigure}{.45\textwidth}
        \centering
         \caption{2 QPUs with 256 data qubits}
        \includegraphics[width=1.\textwidth]{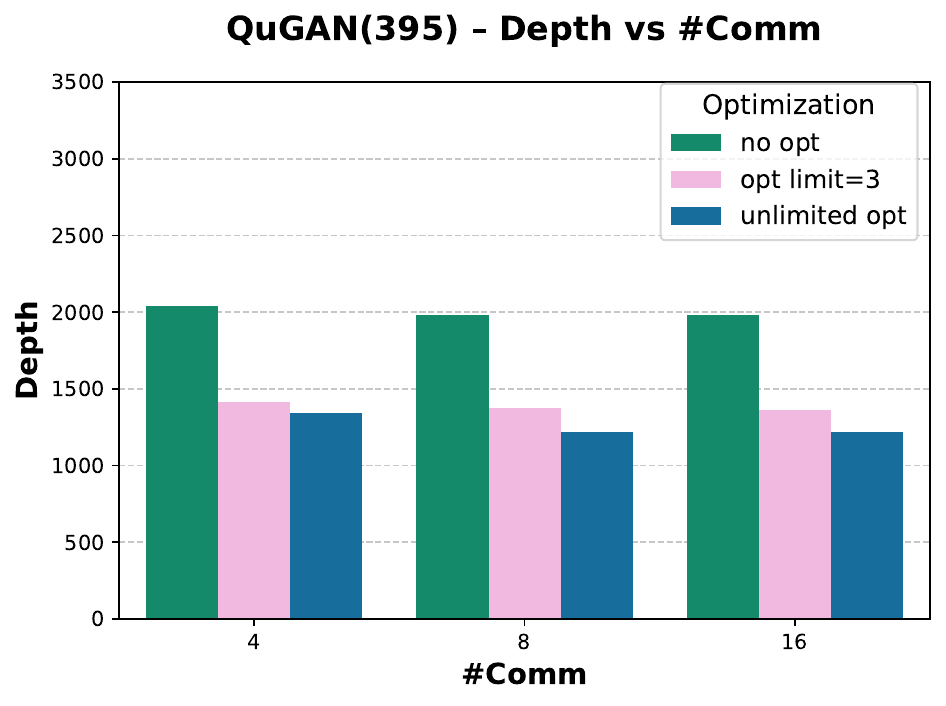}
        \label{fig:QuGAN_2_Depth_vs_Comm}
     \end{subfigure}
     \hfill
     \begin{subfigure}{0.45\textwidth}
        \centering
        \caption{4 QPUs with 128 data qubits}
        \includegraphics[width=1.\textwidth]{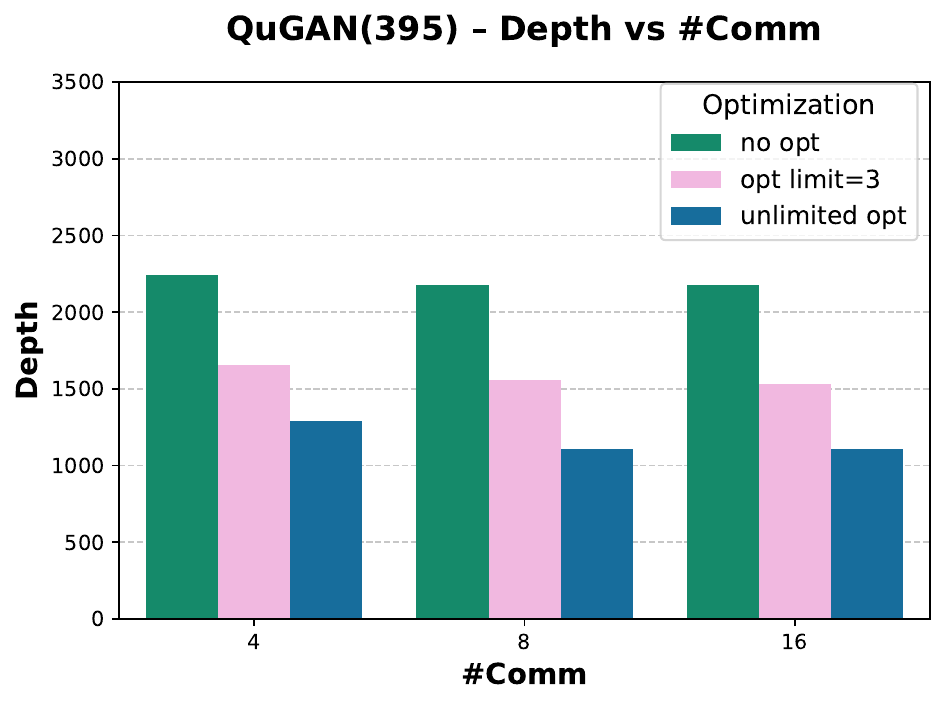}
        \label{fig:QuGAN_4_Depth_vs_Comm}
     \end{subfigure}
     \\
     \begin{subfigure}{0.45\textwidth}
        \centering
        \caption{8 QPUs with 64 data qubits}
        \includegraphics[width=1.\textwidth]{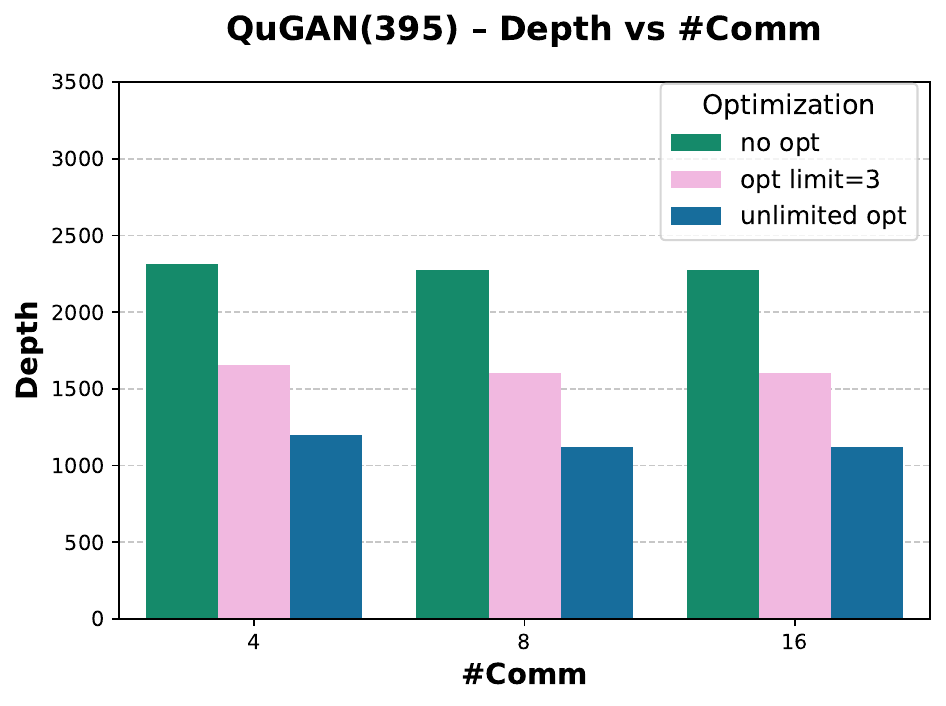}
        \label{fig:QuGAN_8_Depth_vs_Comm}
     \end{subfigure}
     \hfill
     \begin{subfigure}{0.45\textwidth}
        \centering
        \caption{16 QPUs with 32 data qubits}
        \includegraphics[width=1.\textwidth]{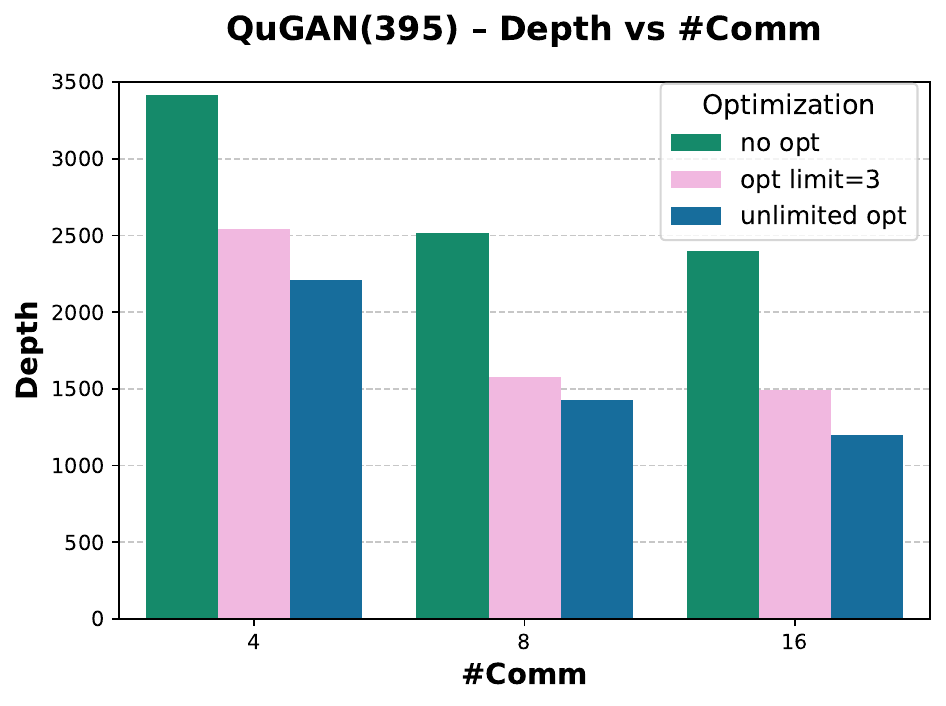}
        \label{fig:QuGAN_16_Depth_vs_Comm}
     \end{subfigure}     
    \caption{Depth of the compiled circuit (QuGAN with 395 qubits) for increasing number of communication qubits for each node. each plot correspnds to a specific DQC architecture: (a) 2 nodes with 256 data qubits, (b) 4 nodes with 128 data qubits, (c) 8 nodes with 64 data qubits, and (d) 16 nodes with 32 data qubits.}
    \label{fig:QuGAN395-Depth_vs_Comm}
    \hrulefill
\end{figure*}

\begin{figure*}
     \centering
     \begin{subfigure}{.45\textwidth}
        \centering
        \caption{2 QPUs with 256 data qubits}
        \includegraphics[width=1.\textwidth]{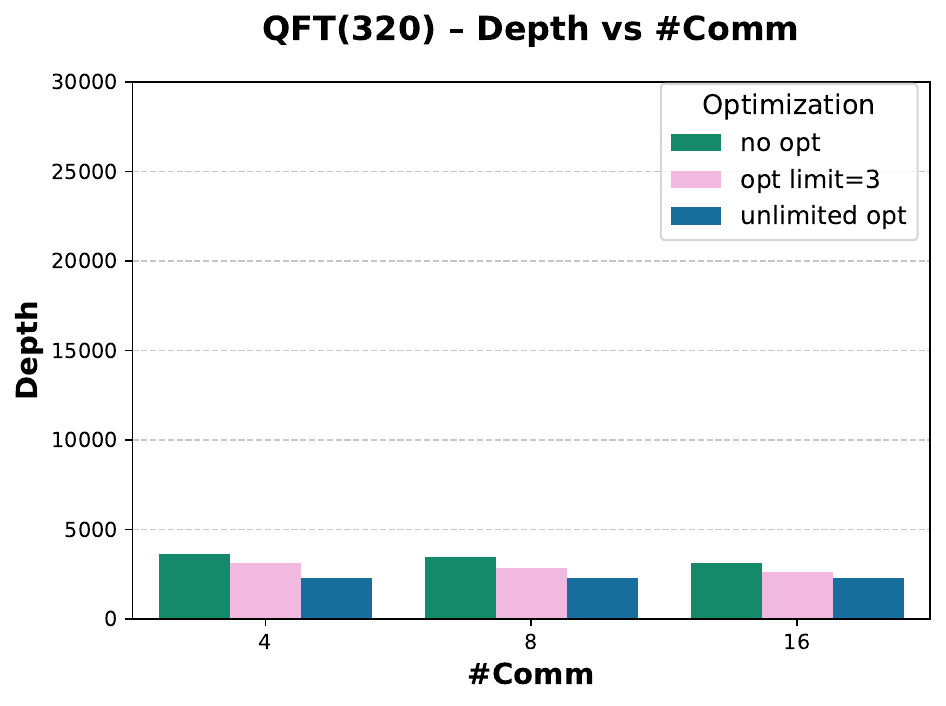}
        \label{fig:QFT_2_Depth_vs_Comm}
     \end{subfigure}
     \hfill
     \begin{subfigure}{0.45\textwidth}
        \centering
        \caption{4 QPUs with 128 data qubits}
        \includegraphics[width=1.\textwidth]{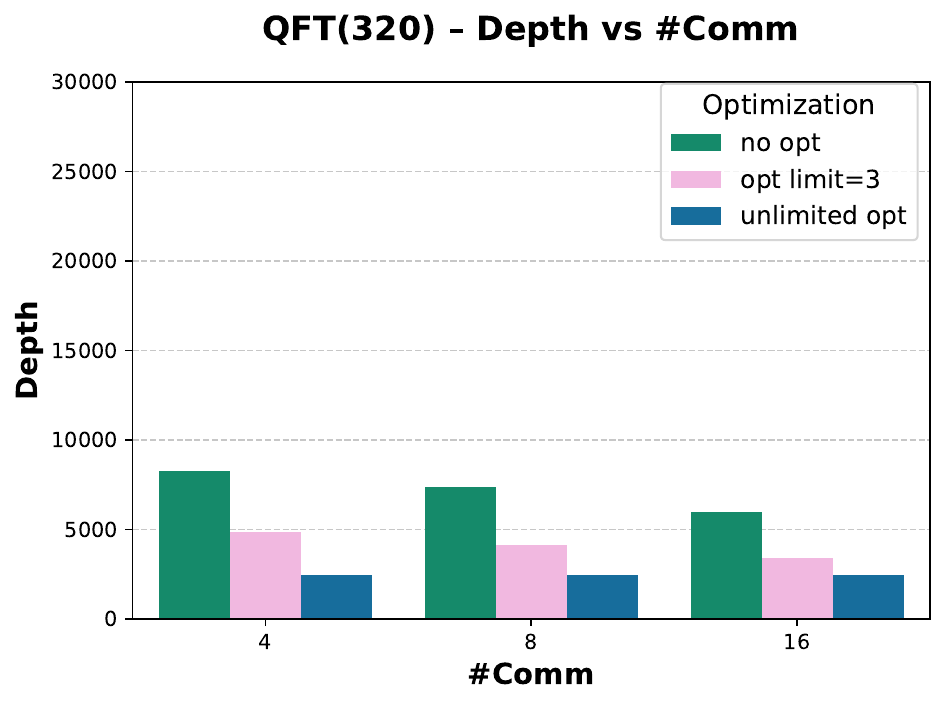}
        \label{fig:QFT_4_Depth_vs_Comm}
     \end{subfigure}
     \\
     \begin{subfigure}{0.45\textwidth}
        \centering
        \caption{8 QPUs with 64 data qubits}
        \includegraphics[width=1.\textwidth]{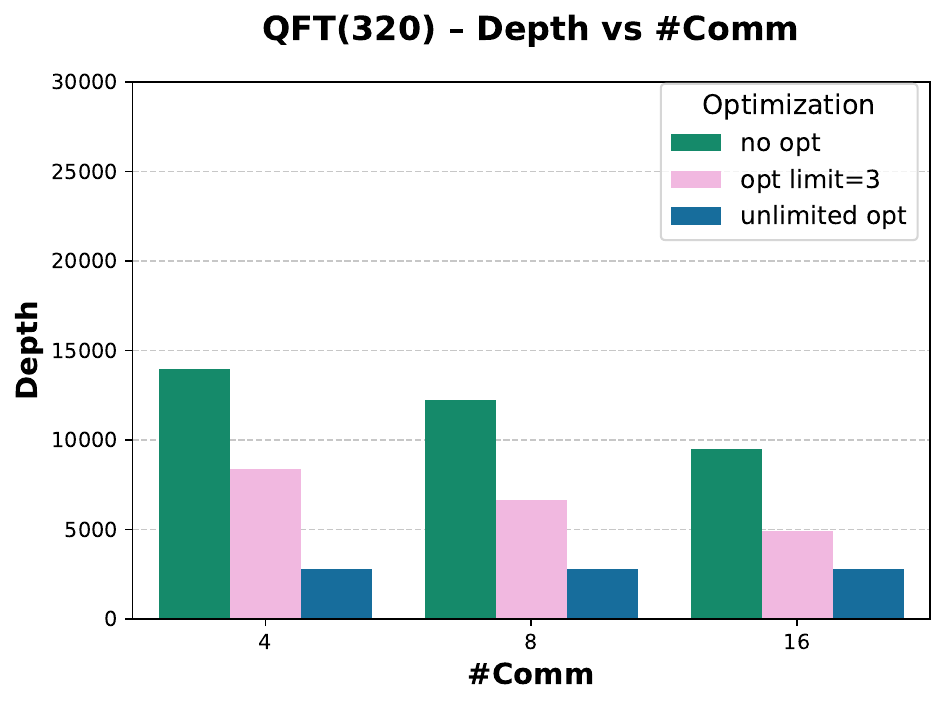}
        \label{fig:QFT_8_Depth_vs_Comm}
     \end{subfigure}
     \hfill
     \begin{subfigure}{0.45\textwidth}
        \centering
        \caption{16 QPUs with 32 data qubits}
        \includegraphics[width=1.\textwidth]{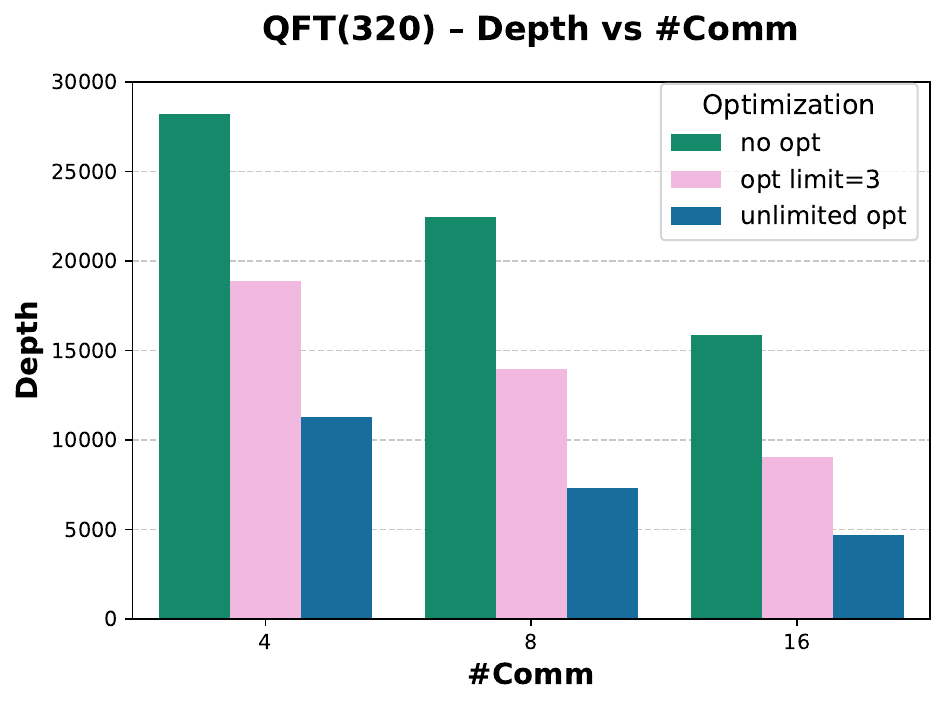}
        \label{fig:QFT_16_Depth_vs_Comm}
     \end{subfigure}     
    \caption{Depth of the compiled circuit (QFT with 320 qubits) for increasing number of communication qubits for each node. each plot correspnds to a specific DQC architecture: (a) 2 nodes with 256 data qubits, (b) 4 nodes with 128 data qubits, (c) 8 nodes with 64 data qubits, and (d) 16 nodes with 32 data qubits.}
    \label{fig:QFT320-Depth_vs_Comm}
    \hrulefill
\end{figure*}

\begin{figure*}
     \centering
     \begin{subfigure}{.45\textwidth}
        \centering
        \caption{2 QPUs with 256 data qubits}
        \includegraphics[width=1.\textwidth]{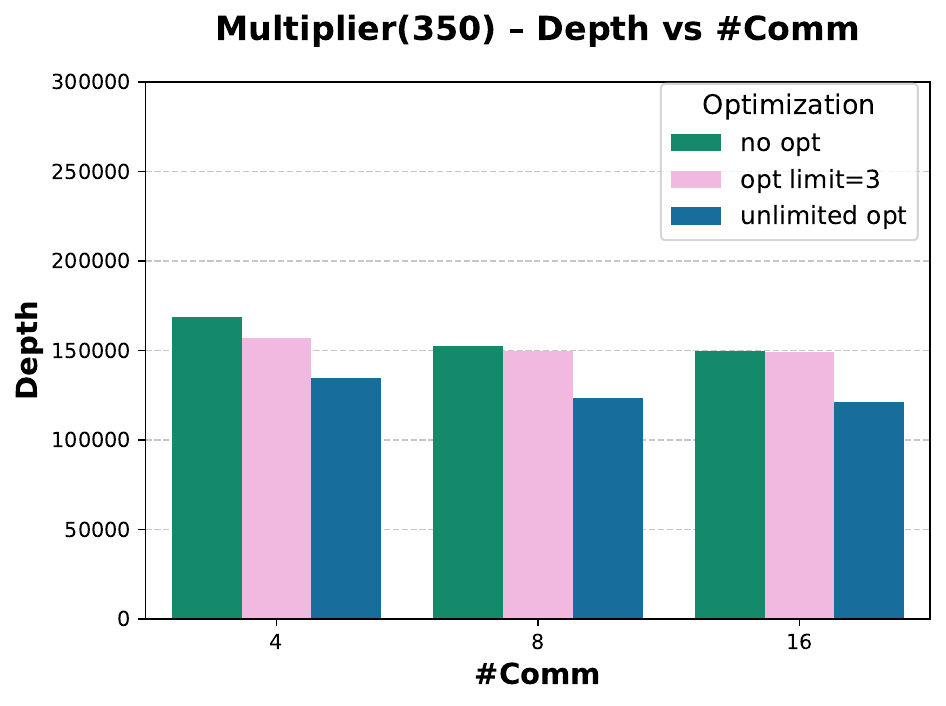}
        \label{fig:Multiplier_2_Depth_vs_Comm}
     \end{subfigure}
     \hfill
     \begin{subfigure}{0.45\textwidth}
        \centering
        \caption{4 QPUs with 128 data qubits}
        \includegraphics[width=1.\textwidth]{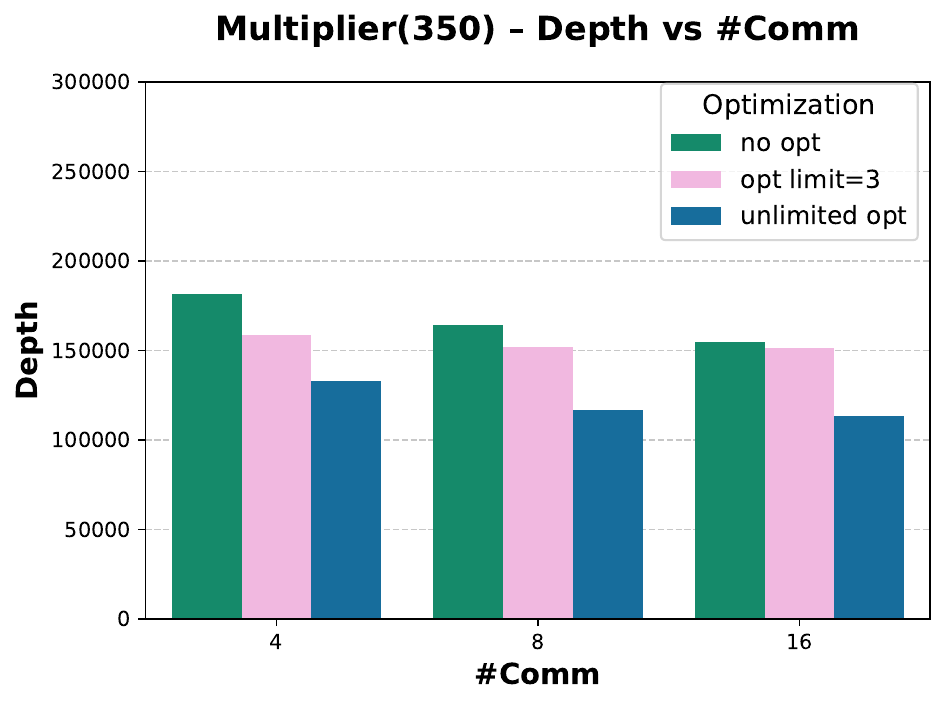}
        \label{fig:Multiplier_4_Depth_vs_Comm}
     \end{subfigure}
     \\
     \begin{subfigure}{0.45\textwidth}
        \centering
        \caption{8 QPUs with 64 data qubits}
        \includegraphics[width=1.\textwidth]{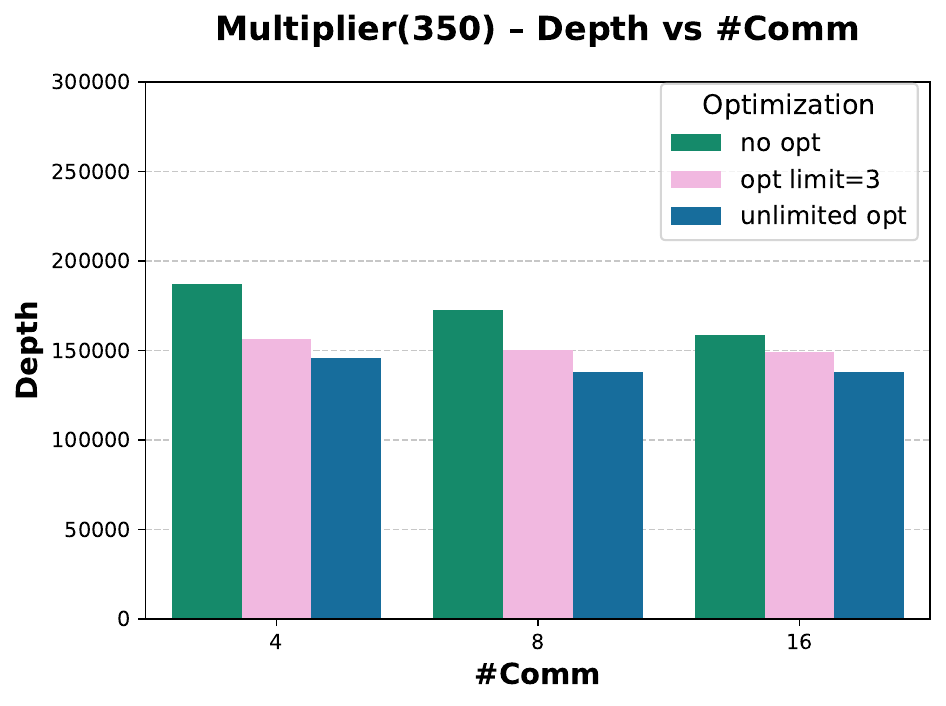}
        \label{fig:Multiplier_8_Depth_vs_Comm}
     \end{subfigure}
     \hfill
     \begin{subfigure}{0.45\textwidth}
        \centering
        \caption{16 QPUs with 32 data qubits}
        \includegraphics[width=1.\textwidth]{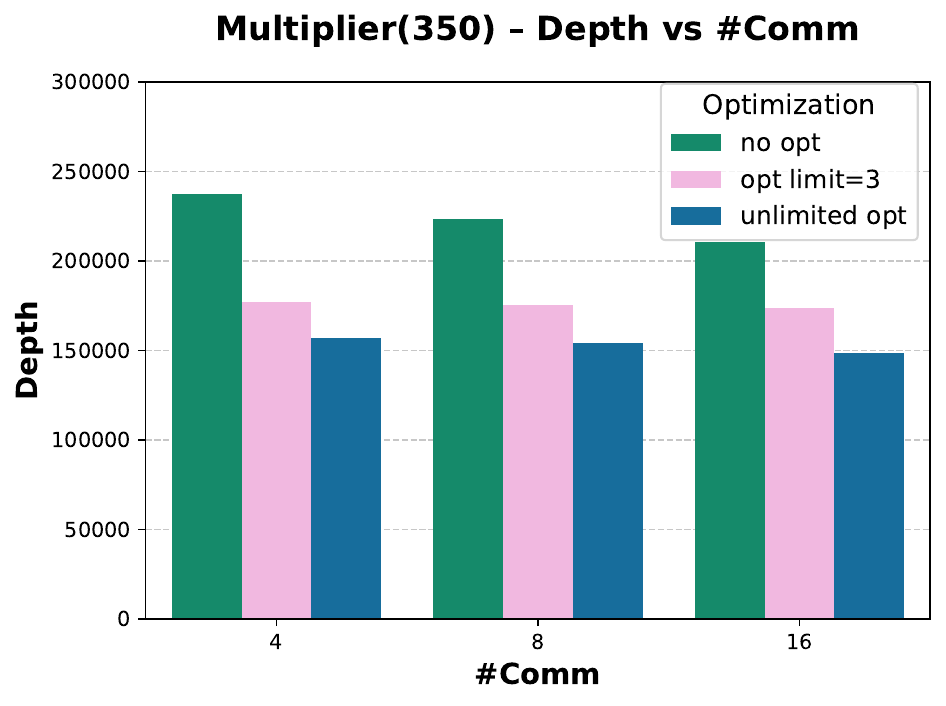}
        \label{fig:Multiplier_16_Depth_vs_Comm}
     \end{subfigure}     
    \caption{Depth of the compiled circuit (Multiplier with 350 qubits) for increasing number of communication qubits for each node. each plot correspnds to a specific DQC architecture: (a) 2 nodes with 256 data qubits, (b) 4 nodes with 128 data qubits, (c) 8 nodes with 64 data qubits, and (d) 16 nodes with 32 data qubits.}
    \label{fig:Multiplier350-Depth_vs_Comm}
    \hrulefill
\end{figure*}

For the QuGAN circuit with 395 qubits, Figure \ref{fig:QuGAN395-Depth_vs_Comm} shows the depth of the compiled circuit, considering devices with an increasing number of communication qubits (from 4 to 8 to 16) and four DQC architectures with an increasing number of QPUs and a decreasing number of data qubits. It may be observed that increasing the number of communication qubits is usually beneficial for reducing the depth of the compiled circuit. This benefit becomes more evident as the number of partitions increases. For the QFT circuit with 320 qubits and for the Multiplier circuit with 350 qubits, the same analysis is reported in Figure \ref{fig:QFT320-Depth_vs_Comm} and Figure \ref{fig:Multiplier350-Depth_vs_Comm}. Especially in the QFT case, the benefit of the increasing number of communication qubits is glaring.

Another relevant insight is the observation that the aforementioned metrics (the number of EPR pairs and the depth of the output circuit) usually worsen when the number of QPUs is significantly larger than necessary. For example, having QPUs with 128 data qubits, it is sufficient to use 3 of them to execute the QFT(320) circuit, resulting in 102 EPR pairs and a depth of 2389 with the best possible optimization. Using 4 QPUs would result in 153 EPR pairs and a depth of 2467. Having QPUs with 64 qubits, it is sufficient to use 5 (resulting in 204 EPR pairs and a depth of 2545). Using 8 QPUs results in 357 EPR pairs and a depth of 2779. Similarly, using 10 QPUs with 32 data qubits is sufficient and more efficient than using 16. However, there are some exceptions. For example, considering the QuGAN(395) circuit and QPUs with 64 data qubits, 7 QPUs are sufficient but less efficient than 8 QPUs in terms of depth (the number of EPR pairs is the same). All the data are reported in Table \ref{tab:results2}.

\begin{table*}[]
\begin{tiny}
\begin{tabular}{ccccccccccc}
Circuit    & $n_q$ & \# QPUs & QPU type     & Input Depth & Output Depth (no opt) & Output Depth (limit=3 opt) & Output Depth (unlimited opt) & EPR pairs (no opt) & EPR pairs (limit=3 opt) & EPR pairs (unlimited opt) \\
\hline
Adder      & 118   & 2       & grid\_64\_4  & 132         & 587                   & 590                        & 588                          & 11                 & 7                       & 5                         \\
Adder      & 118   & 4       & grid\_32\_4  & 132         & 589                   & 588                        & 586                          & 13                 & 9                       & 7                         \\
Adder      & 118   & 8       & grid\_16\_4  & 132         & 647                   & 608                        & 597                          & 67                 & 38                      & 28                        \\
Adder      & 118   & 16      & grid\_8\_4   & 132         & 915                   & 661                        & 643                          & 145                & 113                     & 91                        \\
\hline
QFT        & 320   & 2       & grid\_256\_4 & 2550        & 3651                  & 3108                       & 2311                         & 702                & 364                     & 51                        \\
QFT        & 320   & 3       & grid\_128\_4 & 2550        & 5069                  & 3983                       & 2389                         & 1404               & 728                     & 102                       \\
QFT        & 320   & 4       & grid\_128\_4 & 2550        & 8273                  & 4858                       & 2467                         & 2106               & 1092                    & 153                       \\
QFT        & 320   & 5       & grid\_64\_4  & 2550        & 8798                  & 5733                       & 2545                         & 2808               & 1456                    & 204                       \\
QFT        & 320   & 8       & grid\_64\_4  & 2550        & 13945                 & 8358                       & 2779                         & 4914               & 2548                    & 357                       \\
QFT        & 320   & 10      & grid\_32\_4  & 2550        & 15888                 & 10108                      & 2935                         & 6318               & 3276                    & 459                       \\
QFT        & 320   & 16      & grid\_32\_4  & 2550        & 28206                 & 18870                      & 11284                        & 10272              & 6119                    & 2897                      \\
\hline
Multiplier & 350   & 2       & grid\_256\_4 & 29193       & 168715                & 157047                     & 134751                       & 13228              & 11108                   & 8761                      \\
Multiplier & 350   & 3       & grid\_128\_4 & 29193       & 179932                & 157073                     & 144289                       & 21084              & 15103                   & 12302                     \\
Multiplier & 350   & 4       & grid\_128\_4 & 29193       & 181460                & 158834                     & 132837                       & 23348              & 16857                   & 13320                     \\
Multiplier & 350   & 6       & grid\_64\_4  & 29193       & 180657                & 159773                     & 147882                       & 25906              & 19220                   & 16523                     \\
Multiplier & 350   & 8       & grid\_64\_4  & 29193       & 187353                & 156169                     & 145806                       & 29854              & 21223                   & 17955                     \\
Multiplier & 350   & 11      & grid\_32\_4  & 29193       & 210719                & 167095                     & 156972                       & 48960              & 31174                   & 24269                     \\
Multiplier & 350   & 16      & grid\_32\_4  & 29193       & 237295           & 176856                     & 156865                       & 76372        & 51162                   & 45531                     \\
\hline
QuGAN      & 395   & 2       & grid\_256\_4 & 396         & 2041                  & 1412                       & 1344                         & 408                & 219                     & 29                        \\
QuGAN      & 395   & 4       & grid\_128\_4 & 396         & 2240                  & 1656                       & 1290                         & 612                & 328                     & 42                        \\
QuGAN      & 395   & 7       & grid\_64\_4  & 396         & 2979                  & 1745                       & 1459                         & 720                & 388                     & 67                        \\
QuGAN      & 395   & 8       & grid\_64\_4  & 396         & 2316                  & 1658                       & 1201                         & 724                & 393                     & 62                        \\
QuGAN      & 395   & 13      & grid\_32\_4  & 396         & 3252                  & 2426                       & 2038                         & 889                & 593                     & 291                       \\
QuGAN      & 395   & 16      & grid\_32\_4  & 396         & 3415                  & 2542                       & 2208                         & 1020               & 666                     & 409                       \\
\hline
Swap Test  & 83    & 2       & grid\_64\_4  & 44          & 415                   & 255                        & 264                          & 80                 & 40                      & 4                         \\
Swap Test  & 83    & 4       & grid\_32\_4  & 44          & 455                   & 215                        & 229                          & 120                & 60                      & 15                        \\
Swap Test  & 83    & 8       & grid\_16\_4  & 44          & 479                   & 195                        & 204                          & 144                & 70                      & 28                        \\
Swap Test  & 83    & 16      & grid\_8\_4   & 44          & 487                   & 271                        & 216                          & 152                & 79                      & 40                        \\                
\hline
\end{tabular}
\end{tiny}
\caption{Compilation results for selected circuits, considering different DQC architectures. All QPUs have 4 communication qubits.}
\label{tab:results2}
\end{table*}

\section{Conclusions}
\label{sec:conclusions}

In this work, a novel compiler for DQC is presented that integrates a non-local gate grouping pass, a non-local gate reordering pass, and an improved non-local scheduling pass. To this end, a greedy algorithm explores the circuit and groups together the gates that could share an EPR pair while also changing the order of commutative gates when necessary. In this way, the compiled circuits show reduced depth and EPR usage. 
To take into account that the quality of EPR pairs quickly deteriorates, the number of non-local gates using the same EPR pair can be bounded. Therefore, depending on the features of the target quantum network, the user can achieve different levels of optimization. 
The experimental results show that the proposed approach brings benefits even when assuming a low EPR pair lifetime. 

Regarding future work, there are several interesting directions. The compiler can be further tested with sparse network topologies and specific QPU designs, thus leveraging the local mapping pass and the local routing pass. Alternative qubit assignment passes could be introduced -- for example, to cope with heterogeneous DQC systems. A major foreseen development is the introduction of support for high-dimensional quantum systems (qudits) \cite{Cozzolino2019,Fischer2023,Chiesa2024}, paving the way for fault-tolerant DQC.

\section*{Data Availability}
All data and code required to reproduce all plots presented in this work have been made available at \url{https://zenodo.org/records/18789206}.

\section*{Acknowledgment}
This work was funded by the European Union's Horizon Europe research and innovation programme under grant agreement No. 101102140 – QIA Phase 1. Furthermore, this work benefited from the High Performance Computing facility at the University of Parma, Italy.

\bibliographystyle{unsrt}
\bibliography{references.bib}

\begin{small}

\begin{wrapfigure}{l}{25mm}
\includegraphics[width=1in,height=1.25in,clip,keepaspectratio]{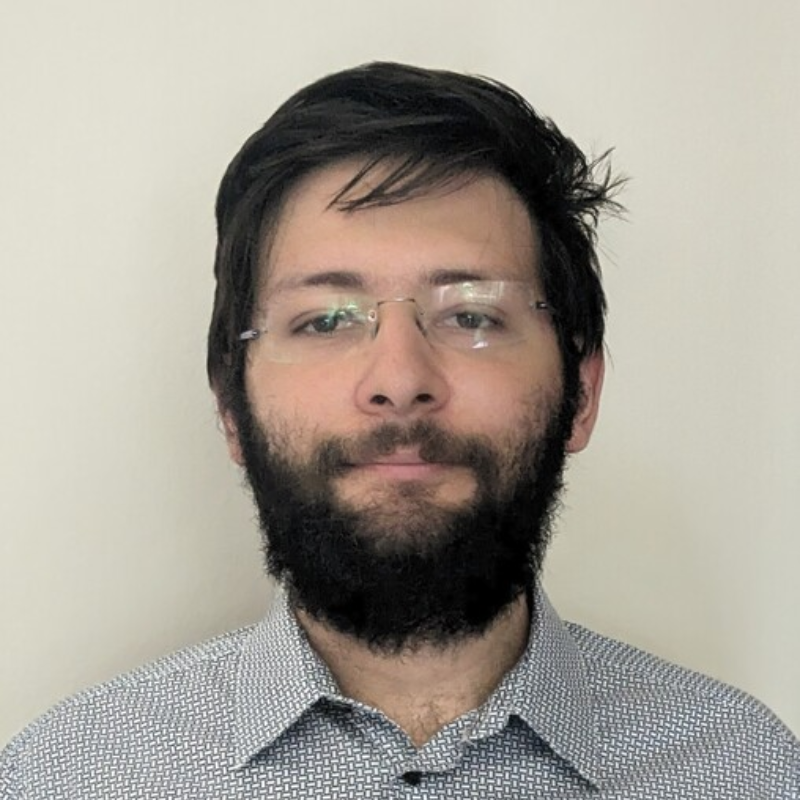}
\end{wrapfigure}\par
  \textbf{Michele Bandini}  received the B.S. and M.S. degrees in Mathematics from the University of Bologna, Bologna, Italy. He is currently a Ph.D. student in Information Technologies at the University of Parma, Parma, Italy. His research focuses on the optimization of resources in quantum compiling for DQC. His work is carried out in the context of the European Quantum Flagship project Quantum Internet Alliance.\par

\vspace{0.2cm}

\begin{wrapfigure}{l}{25mm}
\includegraphics[width=1in,height=1.25in,clip,keepaspectratio]{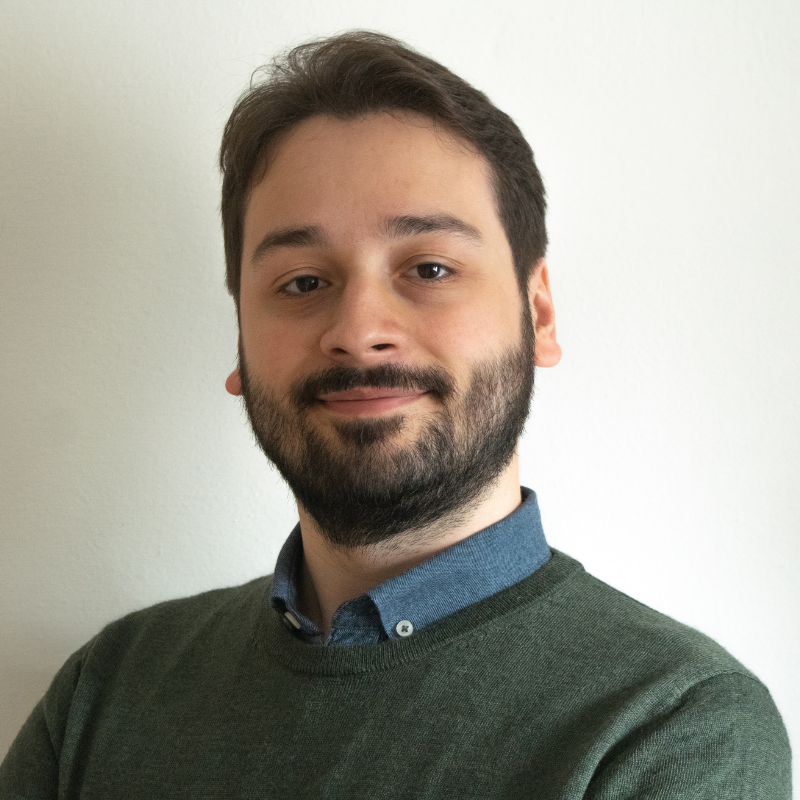}
\end{wrapfigure}\par
  \textbf{Davide Ferrari} (GSN'19-M'23) received his Ph.D. in Information Technologies at the Department of Engineering and Architecture of the University of Parma, Italy, in 2023. During the Ph.D. he worked on quantum compiling, quantum optimization and distributed quantum computing. He has been a research scholar at Future Technology Lab of the University of Parma, working on the design of efficient algorithms for quantum compiling. He is now a research fellow at the Department of Engineering and Architecture of the University of Parma. He is involved in the Quantum Information Science (QIS) research initiative at the University of Parma, where he is a member of the Quantum Software Laboratory. In 2020, he won the 'IBM Quantum Awards Circuit Optimization Developer Challenge'. His research focuses on quantum optimization applications and efficient quantum compiling for local and distributed quantum computing.\par

\vspace{0.2cm}

\begin{wrapfigure}{l}{25mm}
\includegraphics[width=1in,height=1.25in,clip,keepaspectratio]{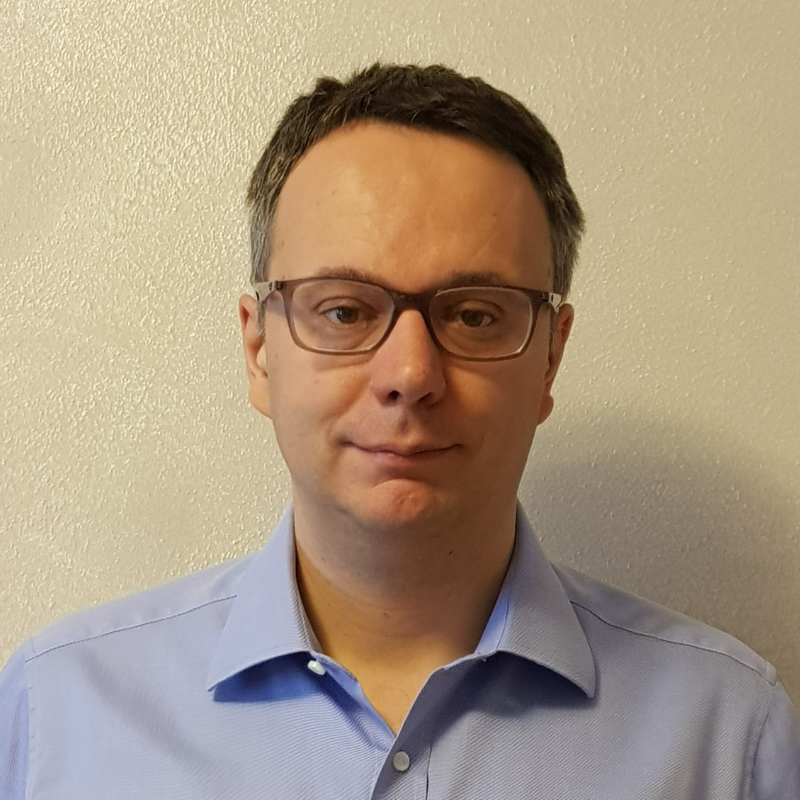}
\end{wrapfigure}\par
    \textbf{Stefano Carretta} received his PhD in Physics from the University of Parma (Italy) in 2005 and he is Full Professor in Theoretical Physics of Matter at the University of Parma. His research activity is mainly focused on the theoretical modeling of the quantum behavior of magnetic molecules and in quantum information processing. He contributed to put forward some of the first proposals for the use of magnetic molecules as qubits and the first proposal and experimental demonstration of the use of molecular nanomagnets as quantum simulators. He was member of the commission of experts on Quantum Technologies for the 2021-27 Italian National Research Program (PNR) and he was involved into several national and European projects involving Quantum Technologies. He is currently one of the PI of an European ERC Synergy Project. He is coauthor of about 180 research papers published in international journals. In 2006 he has been appointed of the "Le Scienze" (the Italian version of Scientific American) medal and of the President of the Italian Republic medal for his research on molecular nanomagnets. In 2011 he has been appointed of the prestigious Olivier Kahn International Award for his contribution to the theory of molecular magnetism.\par

\vspace{0.2cm}

\begin{wrapfigure}{l}{25mm}
\includegraphics[width=1in,height=1.25in,clip,keepaspectratio]{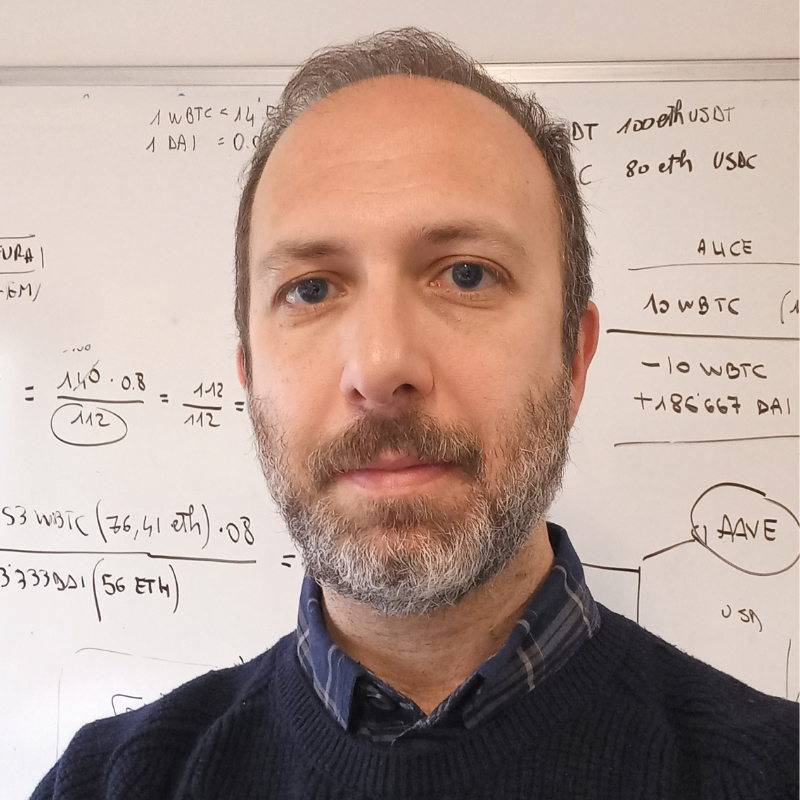}
\end{wrapfigure}\par
    \textbf{Michele Amoretti} (S'01-M'06-SM'19) received his PhD in Information Technologies in 2006 from the University of Parma, Parma, Italy.
    He is Associate Professor of Computer Engineering at the University of Parma, where he leads the Quantum Software Laboratory. In 2013, he was a Visiting Researcher at LIG Lab, in Grenoble, France. His current research interests are mainly in Quantum Computing and High Performance Computing. He authored or co-authored over 150 research papers in refereed international journals, conference proceedings, and books. He serves as \textit{Associate Editor} for the journal IEEE Trans. on Quantum Engineering. He is the Use Case Team Lead of the European HE project Quantum Internet Alliance. He is the CINI Consortium delegate in the UNI/CT 535 ``Quantum Technologies'' UNINFO Commission, which is the national mirror of the CEN/CENELEC JTC 22 ``Quantum Technologies'' standardization committee. He is also a member of the IEC/ISO JTC 3 ``Quantum Technologies'' standardization committee.\par

\vspace{0.2cm}

\end{small}

\end{document}